\documentclass[preprint2]{aastex}

\begin{document}

\title{870 micron observations of nearby 3CRR radio galaxies}

\author{A.~C.~Quillen$^{1,2,3,4}$, Jessica Almog$^{1,5}$,
\& Mihoko Yukita$^{2,6,7}$ 
}
\altaffiltext{1}{Department of Physics and Astronomy,
University of Rochester, Rochester, NY 14627;}
\altaffiltext{2}{Steward Observatory, The University of Arizona, Tucson, AZ 85721;}
\altaffiltext{3}{Visiting Astronomer at the Infrared Telescope
Facility, which is operated by the University of
Hawaii under Cooperative Agreement no. NCC
5-538 with the National Aeronautics and Space
Administration, Office of Space Science,
Planetary Astronomy Program.} 
\altaffiltext{4}{\it aquillen@pas.rochester.edu}
\altaffiltext{5}{\it jessica\_almog@hotmail.com}
\altaffiltext{6}{Harvard-Smithsonian Center for Astrophysics, 
60 Garden Street, Cambridge, MA 02138}
\altaffiltext{7}{\it myukita@cfa.harvard.edu}

\begin{abstract}
We present submillimeter continuum observations at 870 microns 
of the cores of low redshift 3CRR radio galaxies, observed 
at the Heinrich Hertz Submillimeter Telescope.
The cores are nearly flat spectrum between the radio and submillimeter
which implies that the submillimeter continuum is likely to be synchrotron
emission and not thermal emission from dust.   
The emitted power from nuclei detected at optical wavelengths and in the 
X-rays is similar in the submillimeter, optical and X-rays.
The submillimeter to optical and X-ray power ratios
suggest that most of these sources resemble misdirected 
BL~Lac type objects with synchrotron emission peaking at low energies.
However we find three exceptions, the FR~I galaxy 3C264 and
the FR~II galaxies 3C390.3 and 3C338 with high X-ray to submillimeter
luminosity ratios. These three objects are candidate misdirected
high or intermediate energy peaked BL~Lac type objects.
With additional infrared observations and
from archival data, we compile spectral energy distributions (SEDs)
for a subset of these objects.
The steep dips observed near the optical wavelengths in many
of these objects suggest that
extinction inhibits the detection and reduces the flux 
of optical continuum core counterparts.  
High resolution near or mid-infrared
imaging may provide better measurements of the underlying
synchrotron emission peak.


\end{abstract}


\section{Introduction}

A snapshot survey of optical counterparts to 3CR radio sources 
with the Hubble Space Telescope (HST) Wide Field and Planetary 
Camera 2 (WFPC2) has detected unresolved ($<0.1''$)
optical nuclear components to these luminous radio sources
in a large number of objects, particularly at low redshift 
\citep{martel99,zirbel}.  
A correlation between radio core emission at 5GHz and optical core
flux was discovered by \citep{C99,C00,HW00}, and
between the X-ray emission and core radio emission
\citep{HW99,canosa99} implying that 
the X-ray and optical emission are relativistically
beamed and so probably originate in the radio jet.
As pointed out by these studies,
these correlations would be consistent with the unifying model
that has identified populations of FR~I radio sources as BL~Lac type objects
seen at different viewing angles \citep{hardcastle03,T03,bai01,urry95}.

Multiwavelength studies of radio galaxy cores represent a detailed 
way to test the unification model as their spectral energy distributions
(SEDs) can be directly compared to BL~Lac type objects.
Previous work has focused on three flux points (X-ray, radio
and optical) and found that they are not consistent with 
a single power law \citep{HW00}, and could be consistent with the SED
of a BL~Lac type object, but seen at less inclined angles 
with respect to the line of sight \citep{C00}.   By carrying out 
observations of the nuclei in the submillimeter
we extend the coverage of the SED to contain a point intermediate
between the optical and longer wavelength radio emission.
With these additional data we aim to test the hypothesis that the
underlying SEDs are similar to BL~Lac type objects,
estimate the total bolometric output of these
nuclei and search for constraints on the shape of 
the SEDs.

The SEDs of BL~Lac type objects and blazars typically have two energy peaks,
a lower one from synchrotron radiation and a higher one
interpreted to be caused by inverse Compton scattering 
(e.g., \citealt{urry95,ghisellini}).  
The previous classification into ``low-energy'' peaked objects 
(synchrotron peak residing in the infrared; denoted LBL)
and "high-energy" peaked objects 
(synchrotron residing in the UV/X-rays; denoted HBL)
\citep{padovani},
has been superseded by a unified continuum where objects exhibit
a range of synchrotron peak wavelengths 
\citep{fossati, ghisellini,padovani03,beckmann}.

This calls into question the identification of FR~I radio galaxies
with LBL BL~Lac type objects.
Some authors have proposed more complex unification 
scenarios in which the parent
populations of BL~Lac type objects 
include a mix of FR~I and FR~IIs \citep{wurtz}.
Indeed, the extended radio morphologies of BL~Lac objects can
be of both FR~I and II types (e.g., \citealt{kollgaard}).
Thus is more appropriate to unify BL~Lac
objects more generally with radio galaxies \citep{rector}.
By studying the nuclear spectral energy distributions 
of nearby radio galaxies, we aim to find which type
of BL~Lac type object they would resemble if they were oriented with
their jets oriented toward us.

In \S 2 of this paper we present continuum 
observations at 870$\mu$m of the cores
of 34 low redshift radio galaxies.  In \S 3,
we explore the luminosity ratios between our data points and those based
on core fluxes measured at 5GHz, optical and X-ray wavelengths 
compiled by previous studies.
In \S 3 we also compile as detailed SEDs of as many objects as possible
to see if the SEDs do indeed resemble those of BL~Lac type objects and
if so which type.  A discussion follows in \S 4.

\section{Observations}

The 3CRR sample \citep{laing83}
is a flux limited sample of the northern sky.
It includes all objects with 178MHz flux density greater than 
10.9Jy having $\delta > 10^\circ$ and $|b|>10^\circ$ but
excludes the starburst galaxy, M82.  In this paper we
restrict our study to the low redshift component, $z<0.1$, of this sample.
Many of the galaxies in this subsample have been observed at high angular 
resolution  ($\sim 0.1''$) at visible wavelengths
with the Wide Field Planetary Camera 2 (WFPC2) on the Hubble
Space Telescope \citep{martel99} and subsequently in the ultraviolet
with the STIS NUV MAMA detector \citep{allen}.

We used the Heinrich Hertz Submillimeter Telescope
(HHT)\footnote[1]{The HHT is operated by the
Submillimeter Telescope Observatory on behalf of Steward Observatory
and the Max Planck Institut f\"ur Radioastronomie.} \citep{baars}
located on Mt.~Graham, Arizona.
Observations were carried out 2001 Feb 10-12, 
with the 19-channel bolometer array which was
developed by E.~Kreysa and collaborators at the Max-Planck-Institut 
f\"ur Radioastronomie (MPIfR), Bonn \citep{kreysa}.
The 19 channels are located in the center and on
the sides of two concentric regular hexagons, with an apparent spacing
between two adjacent channels (beams) of $50''$. The central frequency
of the bolometer is about 345\,GHz (the highest sensitivity is reached
at 340\,GHz), and the instrument is sensitive mainly between 310 and
380\,GHz. 

To calculate the atmospheric zenith opacity, we made
skydip observations every 40 to 80 minutes.  During the observations
the sky opacity at 345\,GHz was
around 0.4 most of the time, increasing to 0.9 for a few scans.
During the observations, the subreflector
was wobbled at 2\,Hz in azimuth, with a beam throw of $100''$.
Our observation mode consisted of 20 continuum on-off scans, 10 seconds each.  
Except for the brightest sources, we observed each radio galaxy
6 times in the above mode resulting in a total typical on-source exposure time
of 1200 seconds and an actual observing time approximately 3 times this.   

For calibration purposes every few hours we also performed 
mapping and on-off measurements of planets (Venus, Mars).
These measurements
yielded a conversion factor from observed counts to mJy/beam of 
0.8  -- 1.1\,mJy\,beam$^{-1}$\,count$^{-1}$. 
The conversion factor is affected by the atmospheric condition during 
the observations and the uncertainties in the opacity calculation.
The pointing of the telescope was checked every few hours.  Pointing
errors (a few arcseconds) remained well within the beam width 
$\sim 22''\!\!.7$ 
throughout these observations.

Data reduction was performed with 
a customized version of the software package NIC which is 
part of the GILDAS software package and is a collaborative
project of the Centre d'Etudes de Saclay, MPIfR Bonn, the Observatoire
de Grenoble and IRAM (Institut de RadioAstronomie Millim\'etrique).
After baseline subtraction and the elimination of spikes
in each single coverage, the atmospheric noise, which is highly correlated
between the individual channels, was subtracted. The maps were gridded,
restored, and finally combined (with an
appropriate weighting) into a single map per galaxy.
We expected the radio source 
to be located in the central channel of the array, and for
all sources detected this was indeed the case.  
The flux of the central source was measured using an on-off distance
of $120''$ and by subtracting the mean of the sky fluxes from the surrounding
channels.  For most of the galaxies we had multiple observation sets and
the final flux was measured using the noise weighted sum of the flux
of the central source in each map.

Of the 46 galaxies from the 3CRR low redshift galaxies 
observed by \citet{martel99},
we observed 34 galaxies at the HHT.  In addition we include
a measurement of the radio galaxy NGC 6251 which has also been observed
extensively by HST.  NGC 6251 fits the criterion for being selected 
in the 3CRR galaxy \citet{laing83}.
Of the 35 galaxies observed we detected 24 and 
estimated $2 \sigma$ upper limits for the 11 undetected.
The measured fluxes and estimated upper limits are listed in Table 1.
Observations of three galaxies (3C35, 3C198, 3C227) 
were repeated during better observing conditions later in the observing
run,  but remain undetected.  These are listed twice in Table 1.


\section{Results}

\begin{figure*}
\plotone{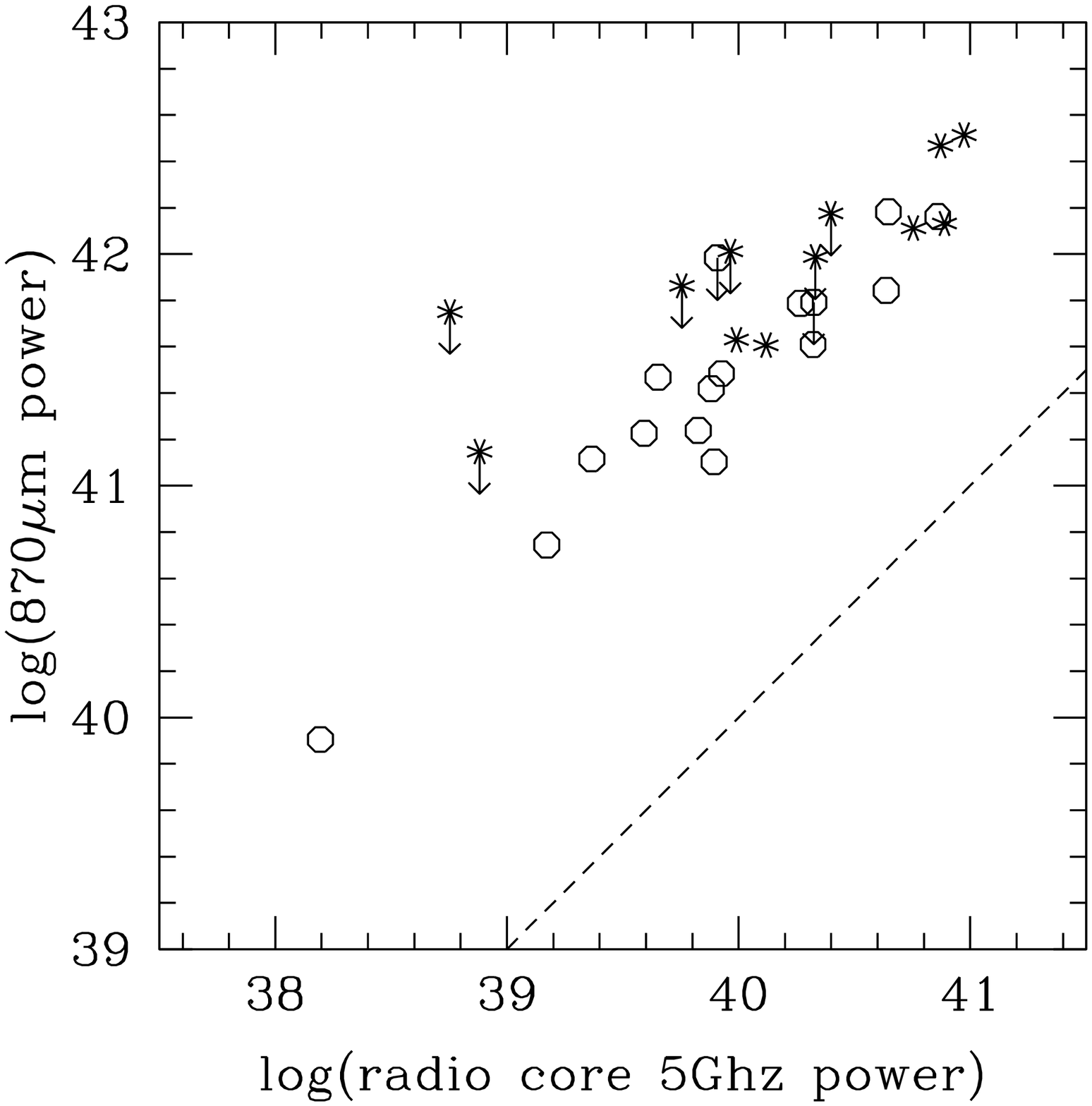}
\plotone{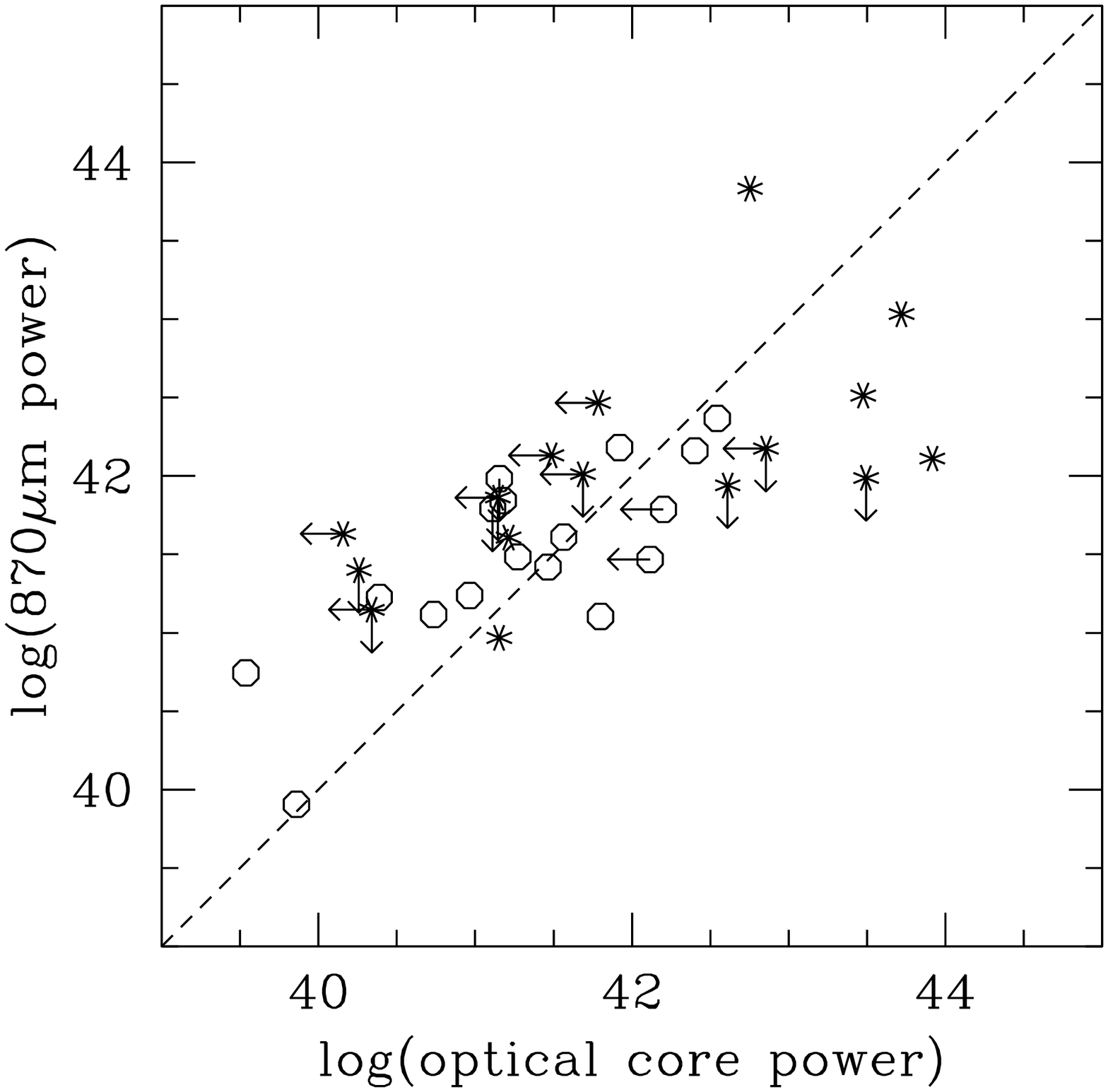}
\plotone{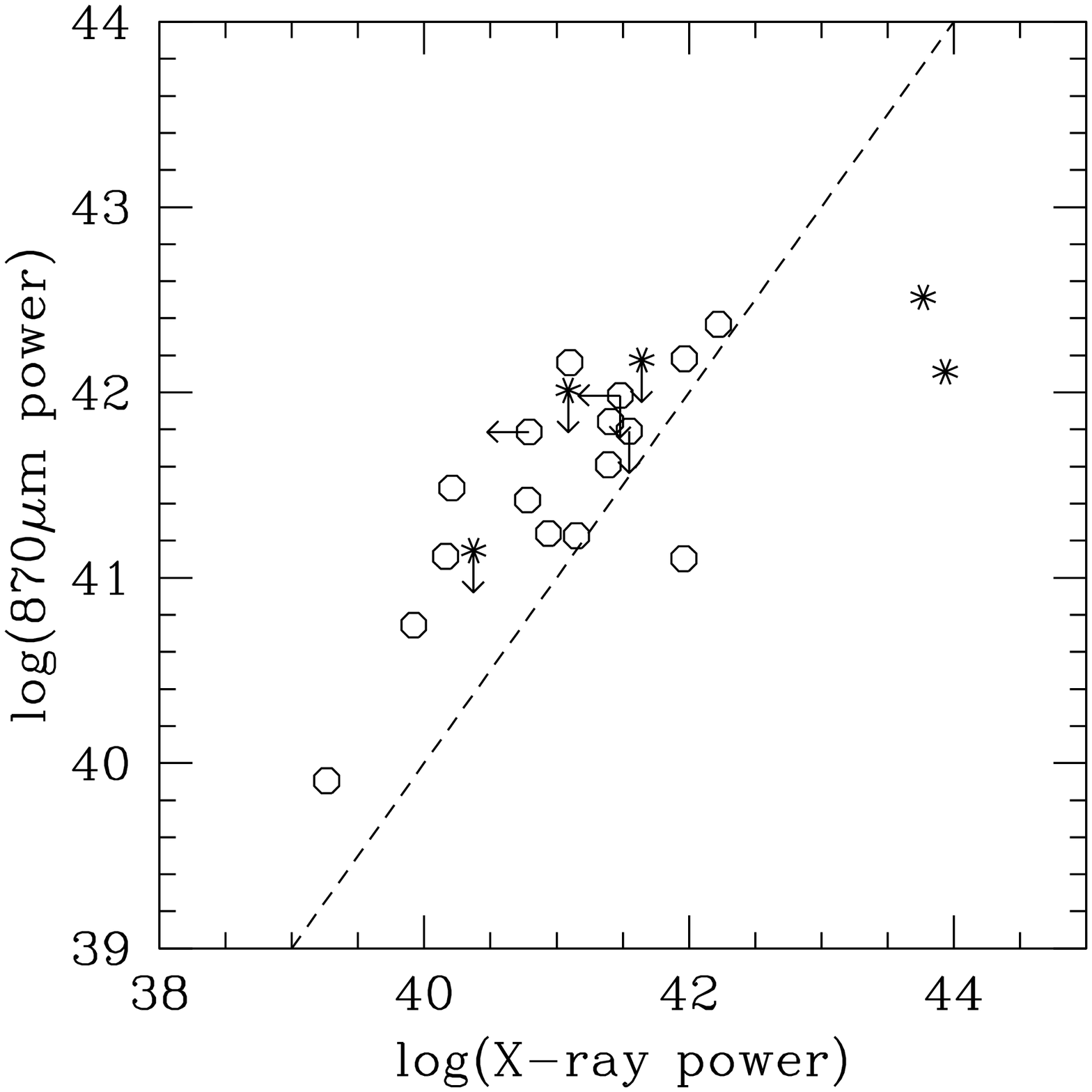}
\caption[junk]{
A comparison of the power emitted at different wavelengths.
a) Comparison between the core power emitted at 5GHz and that 
emitted in the submillimeter at 345\,GHz.
FR~I's are shown as open circles and FR~II's as stars.
The log of the luminosities in ergs~s$^{-1}$, (estimated by $\nu L_\nu$) 
are shown.
The dashed line shows $\alpha=1$ where the flux $S_\nu \propto \nu^{-\alpha}$,
which corresponds to a power $\nu L_\nu$ independent of $\nu$.
The luminosity is the same at the two frequencies for points on this line.
The submillimeter powers are about 50 times larger than those at 5GHz.
Radio 5GHz core fluxes are those measured by \citep{M93,G88}.
b)  Comparison between optical and submillimeter power.
c)  Comparison between X-ray and submillimeter power.
Optical points are taken from those compiled and measured
by \citep{T03,C02b,C99,C00,HW00}, those in the X-rays
by \citep{HW99,HW00,S99,T03,C03,H01}.
As noted by previous studies (e.g., \citealt{C00}), 
some FR~II galaxies have excess optical or X-ray power compared
to their radio power.   Here we find that the
same is true for the optical or X-ray power compared
to that emitted in the submillimeter.
\label{fig:power}
}
\end{figure*}

For many of these galaxies, nuclear continuum fluxes have been measured
in the optical using HST images \citep{T03,C02b,C99,C00,HW00}, in the X-rays
from Chandar, ROSAT and ASCA data
\citep{HW99,HW00,S99,T03,C03,H01}.
Using these measured fluxes and 5GHz core fluxes measured
by \citet{G88,M93}, we
search for correlations and measure spectral indexes
between the radio and submillimeter, optical and X-rays.
Plots comparing measured core powers are shown in Figure \ref{fig:power}.
Luminosities have been  estimated by $\nu f_\nu 4 \pi D^2$ 
where $f_\nu$ is the flux density, $\nu$ the frequency
and $D$ the distance which we estimate from the redshift
using a Hubble constant of 75~km~s$^{-1}$~Mpc$^{-1}$.
Spectral indexes for objects detected at 870$\mu$m are listed
in Table 2 along with luminosities estimated from the submillimeter
fluxes.
The spectral index, $\alpha$, is defined in the sense that 
flux density is proportional to $\nu^{-\alpha}$.

In Figure \ref{fig:power}a we see that there is a correlation
between the $870\mu$m and 5GHz core luminosities.
This implies that the emission at $870\mu$m is not thermal emission
from cold dust but more likely to be synchrotron emission.
Contamination from CO(3-2) line emission in the 345\,GHz band is unlikely.
We find that the typical spectral indexes between
the radio (5GHz core) and submillimeter, $\alpha_{sr} \sim 0.2$
This suggests that
these sources are approximately flat spectrum up to frequencies
higher than 345\,GHz.

There is significantly more power emitted from these sources
in the submillimeter than at 5GHz.  The luminosity ratio 
between the submillimeter and 5GHz is for the mean approximately 50.
This is similar to what is seen in BL~Lac type objects which can have 
their synchrotron emission SED peaking anywhere between the IR through
the X-rays  (e.g., \citealt{fossati,ghi02}).

From Figures \ref{fig:power}b,c
we find that the submillimeter luminosities are similar
to those emitted in the optical and X-rays.
However the scatter in the optical/radio and optical/submillimeter
plots is higher than that seen in the submillimeter/radio
plots.  A large scatter in the optical/submillimeter spectral
index, $\alpha_{os}$, could be caused by a variety of
processes including extinction from dust, an additional
source of radiation such as continuum emission from a broad  
line region, or variations in the position of the emission peaks
from an underlying blazar or BL~Lac type SED.
BL~Lac type objects vary in the wavelength position of their synchrotron peaks,
and the location of these peaks also depend on orientation angle or beaming.
If the underlying SEDs of these sources are similar to blazars or BL~Lac type
objects we might also expect a large scatter in $\alpha_{os}$ 
due to differences in synchrotron peak wavelength and orientation angle.

\begin{figure*}
\epsscale{1.3}
\plotone{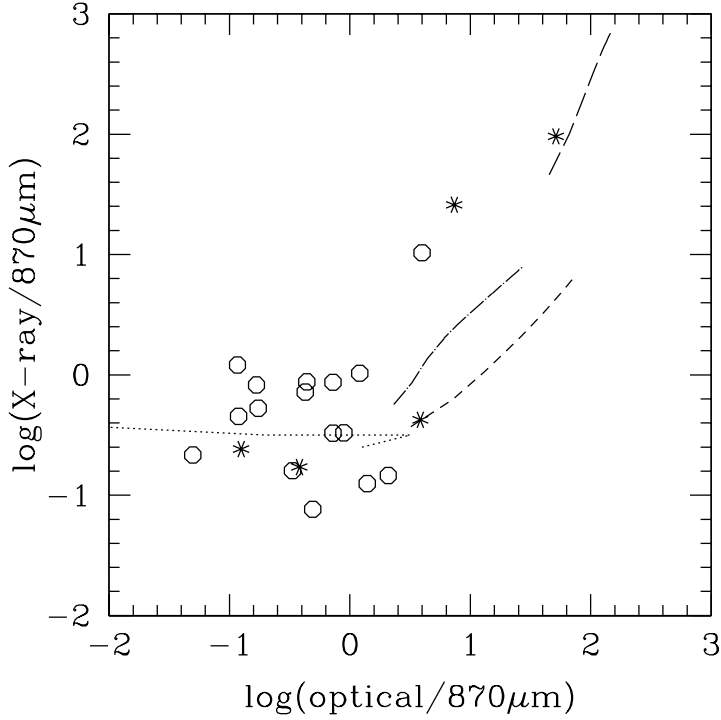}
\caption[junk]{
Comparison between the ratio of power emitted in X-ray to submillimeter
and the ratio of power emitted in optical to submillimeter.
FR~I's are shown as open circles and FR~II's as stars.
The FR~I with the large X-ray/870$\mu$m luminosity ratio is 3C264.
The FR~IIs on the upper right are 3C390.3 and 3C338.
Also shown are the expected ratios for debeamed BL~Lac type objects.
The dotted line shows that expected for an LBL (based on the spectrum
of Mkn421 as done by \citealt{C00}).
The dashed line shows that based on the mean of
the LBL dominated 1Jy radio selected BL~Lac sample measured by 
\citet{fossati} in their Table 5. The leftmost point on 
the line corresponds to that of the observed BL~Lac sample and
the rightmost point in the line  
refers to a debeamed object 
at an orientation angle of $60^\circ$ assuming $\gamma =20$.
The dashed line shows that based on 
the mean of the X-ray selected Einstein Slew sample  (HBL dominated)
as measured by \citet{fossati}.  The dot dashed line shows that
based on the mean of the Wall \& Peacock flat spectrum radio
quasar sample as measured by \citet{fossati}.
The objects 3C264, 3C390.3 and 3C338 are candidate low luminosity HBLs.
The remaining objects are consistent with LBLs 
at a range of beaming angles.  If dust affects the optical luminosity
of the core emission then the true location of the synchrotron components
would be further to the right on this plot.
\label{fig:ratio}
}
\end{figure*}

For most of these sources, the luminosity in the submillimeter
and the X-rays are similar.  \citet{chiaberge,T03}
noted that since the X-ray fluxes are not higher than
the optical fluxes, most of these sources are unlikely to be candidate
debeamed HBL type objects. However extinction could have 
reduced the observed optical fluxes.
The submillimeter observations have the advantage over the optical ones 
that they are relatively unaffected by extinction from dust.
In Figure \ref{fig:ratio} we show the X-ray to submillimeter
and optical to submillimeter luminosity ratios for objects detected
at all three wavelengths.
Most of our sources have similar submillimeter and X-ray luminosities,
confirming the result by \citet{chiaberge} that most of these
sources are more likely to be similar to LBLs rather than HBLs.

In the subsample containing both X-ray and submillimeter measurements, 
we find that only 3 objects have X-ray luminosities significantly
higher than submillimeter
luminosities, the FR~I galaxy, 3C264, and the FR~II galaxies, 
3C382 and 3C390.3.   
These objects could
have SEDs similar to debeamed HBL type objects.  We concur
with \citet{T03} who also found that 3C264 is similar to an HBL.
To test this possibility we have plotted on Figure \ref{fig:ratio} 
luminosity ratios for
debeamed LBL and HBL objects at different orientation angles, based
on Mkn421 (dotted line) as done by \citet{C00} and the means
of the populations measured by \citep{fossati} based on the X-ray
selected Einstein Slew BL~Lac sample (primarily HBLs),
the Wall \& Peacock flat spectrum radio quasar sample and
the 1Jy BL~Lac sample consisting primarily of LBLs.
As these BL~Lac type objects are debeamed (oriented closer to perpendicular to
the line of sight) the optical to submillimeter ratio increases so
the rightmost end of these lines are the objects oriented
nearly perpendicular to the line of sight.

Most of 3CRR sources lie in the region expected for LBLs at a range
of orientation angles,
excepting the three extreme objects 3C264, 3C382 and 3C390.3.
A correction for extinction from dust and photoelectric
absorption in the X-rays would increase the underlying
optical and X-ray fluxes moving the data points upward and
to the right in the plot.  Consequently some of the 
sources with low X-ray to submillimeter luminosity ratios
could be similar to objects intermediate to LBLs and HBLs.   
However the three
extreme galaxies are unlikely to resemble LBLs and so are
candidate HBLs or intermediate objects.

In some cases we have additional information on the object's
orientation.
Most but not all of FR~Is have disks nearly perpendicular to the line
of sight and so are likely to be debeamed compared to BL~Lac type objects.  
Those that contains optical jets 
or exhibit circular dusty disk (such as 3C78 and 3C264) 
are unlikely to have jets oriented perpendicular
to the line of sight \citep{sparks}.

3C264 is probably closer to face on rather than edge-on
because it exhibits a nearly round dusty disk, seen in absorption
at optical wavelengths
\citep{sparks}. 3C390.3 exhibits superluminal
motion \citep{alef96}, and the jets of 3C382 and 3C390.3 are
one sided \citep{black92,leahy95,G01} 
suggesting that these two objects are also oriented at high inclination
angles to the line of sight.
The candidate HBLs we have identified 
are probably not oriented perpendicular to the line of sight.
While 3C390.3 falls near the expected location of the
X-ray selected HBL sample, 3C382 and 3C264 are more likely
to be similar to intermediate type objects, consistent
with recently identified BL~Lac populations with spectra
in between those of LBLs and HBLs
(e.g., \citealt{beckmann,fossati}).

\subsection{Spectral Energy Distributions}

It is clear from the previous section that the possibility
of extinction in the optical bands 
causes an uncertainty which inhibits our study
of the underlying SEDs.  
Because the SEDs of BL~Lac type objects are smooth, the
spectral indexes are only weakly dependent upon
the boosting factor (e.g., \citealt{chiaberge}).
This presents a possible way to tell the difference between
obscuration by dust and beaming.  If the SED is rapidly
changing between the optical and UV wavelengths 
then it is likely that dust is causing significant obscuration resulting
an underestimate of underlying synchrotron
emission in the optical region.

If dust is indeed a factor causing us to underestimate
the extent of the emission in the optical region then
some of these source might be much brighter in the near-IR.
Unfortunately only a few of these galaxies were observed by NICMOS.
\cite{capetti} found that the SED of 3C264 between 1-2 microns
was nearly flat, whereas that of 3C270 decreased remarkably
toward shorter wavelengths.   Here we compile the SEDs of
the objects which were observed at more than one wavelength 
to examine the shape of the SED in more detail
than presented by \cite{capetti, T03}.  
We have compiled in Table 3  core fluxes
in the UV and optical from STIS and WFPC2 images, in
the near-infrared with NICMOS images (on board HST) , in the mid-infrared
from ISOCAM/ISO images and at $3\micron$ from ground based
images taken by us at the IRTF.  We have also included measurements
by and compiled by \citet{C02,C02b,T03}.

\begin{figure*}
\epsscale{1.5}
\plotone{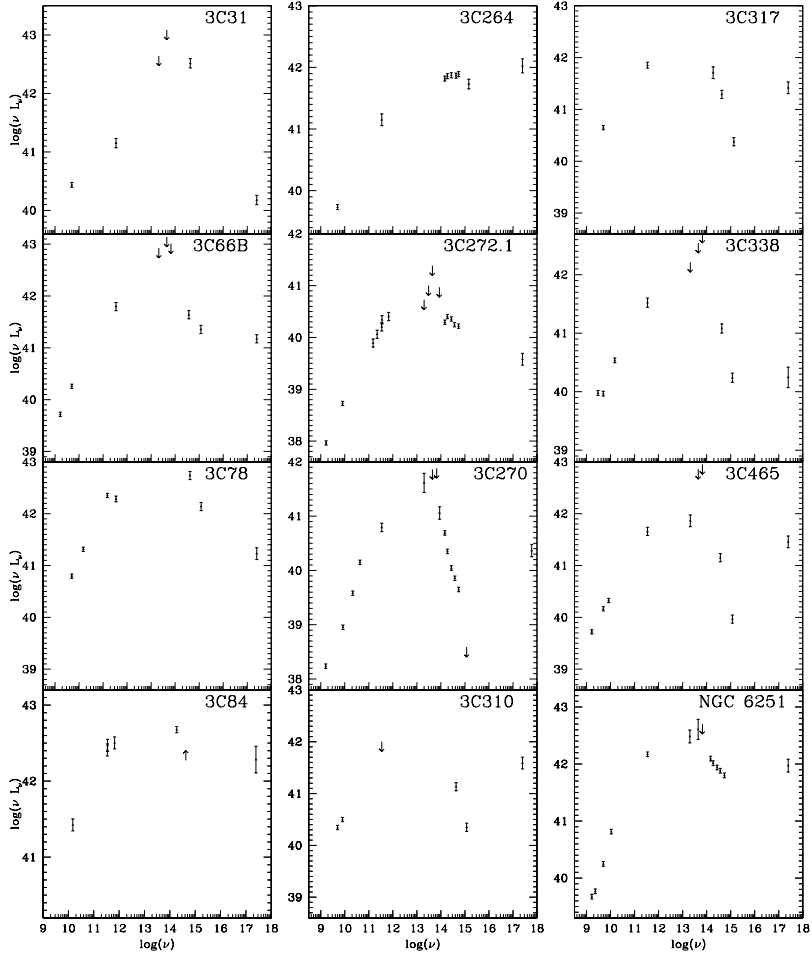}
\caption[junk]{
The SEDs of FR~I nuclei for which we have many measured fluxes.
See the accompanying tables for more information.  
3C270(NGC~4261) and NGC~6251 
have prominent edge-on dusty disks seen in optical
HST images, whereas 3C264 has a face-on one.
3C272.1(M84) has a warped dusty disk.   The steep drop in the SEDs
of many of these galaxies between the near-infrared and UV 
is likely to be caused by extinction from dust.
Excepting that of 3C264, all of these objects have
SEDs resemble LBLs.
We note that much of the scatter among these data is
likely to be caused by intrinsic variability.
\label{fig:fr1}
}
\end{figure*}

\begin{figure*}
\plotone{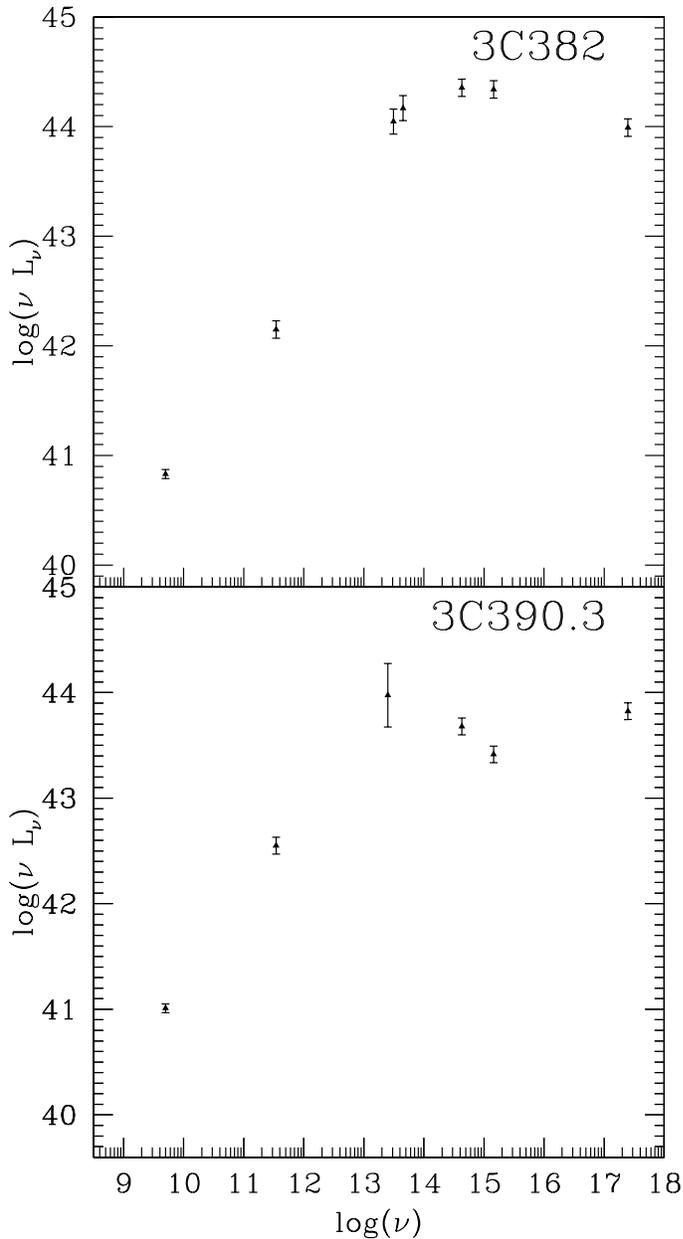}
\caption[junk]{
The SEDs of two FR~II nuclei for which we have many measured fluxes.
See the accompanying tables for more information.
These two objects have high X-ray to submillimeter luminosity ratios
suggesting that they could be similar to HBLs.
From these SEDs, the synchrotron peaks
are most likely to reside in the optical/UV region, intermediate
between those of HBLs and LBLs particularly since 
extinction may have affected the optical/UV observations of 3C390.3.
At an orientation angles pointed toward the observer, the spectrum would
be harder pushing the expected location of the synchrotron peak
toward the UV/Xray region.  Therefor these objects could be candidate
debeamed HBLs.
The large X-ray to submillimeter luminosity ratios suggests that the high level
of optical emission is not due to an extra component of emission
(such as could be from a broad line region or accretion disk)
but is a result of a higher energy peak of the synchrotron emission component.
\label{fig:fr2}
}
\end{figure*}

In Figures \ref{fig:fr1}, and \ref{fig:fr2} 
we show the compiled SEDs for FR~Is and FR~IIs respectively.
These are 
an improvement on previous studies such as \citet{capetti2002,T03} 
which contained fewer data points.
NGC~4261, and NGC~6251, 
have notable nearly edge-on dusty disks seen in HST images,
and they both have extremely steep SEDs between the near-infrared
and optical region, as previous studies have noted \citep{ferrarese,C03}. 
Non-thermal processes such
as synchrotron radiation seldom permit extremely sharp
drops in a spectral energy distribution.  
Variability could cause some of the scatter among the points
in Figures \ref{fig:fr1}, and \ref{fig:fr2}, 
however it would not systematically account for a steep drop
between the near-infrared and UV wavelengths which is seen in many
of these objects.
It would be difficult to construct a model from non-thermal emission 
processes alone in which a number of objects are likely
to exhibit such a steep drop exactly in the near-IR/optical region.
However extinction by dust naturally causes a steep drop in this 
region.  If we assume that the underlying SED
is fairly flat in terms of luminosity, then extinction
causes a reduction in the optical flux by more than
a factor of 10 in NGC~4261 and by a factor of a few in NGC~6251.  
We suspect that extinction from dust
is important even though many of these radio galaxies
were detected as unresolved sources in optical images.

If extinction by dust does affect the detection
of unresolved nuclei in the radio galaxies,  we expect 
that subsequent surveys will
detect more nuclear sources in the infrared than previous detected
in optical surveys.  The energy absorbed by the dust should
be re-emitted in the infrared, so some of the mid-IR emission
should be from hot dust.  If we corrected the optical fluxes
for extinction, the underlying SED should have more optical
emission and consequently
lower optical/X-ray and optical/submillimeter spectral indexes.
The shape of the SEDs displaying sharp
drops between the near-infrared and UV could still be consistent with those of
misdirected LBL type objects.

\section{Summary and Discussion}

In this paper we have presented a survey of submillimeter 
continuum observations at 870 microns
of the low redshift 3CRR radio galaxies, which we observed
with the Heinrich Hertz Submillimeter Telescope.
We find that the submillimeter luminosities are about 50 times
larger than that from the core 5GHz, but similar in size
to the X-ray and optical luminosities.   The SEDs for most of the FR~Is
are similar to what is expected from debeamed
low energy peaked BL~Lac type objects, agreeing with the results
of previous works \citep{chiaberge, T03}.
We predict that the SEDs of most of these objects should peak 
in the infrared and so can be detected with forthcoming mid-IR imaging 
studies.

By examining a few SEDs in detail we have found good evidence
for extinction in the optical region in
sources with edge-on disks (M84, NGC 4261, NGC 6251) 
but not in a source with a face-on disk (3C264).
We suspect that some of the scatter in optical spectral indexes is
due to extinction from dust, and that many of these nuclei should
be brighter at near and mid-infrared wavelengths.

The ratio of submillimeter to X-ray and optical luminosities suggest
that most of our sources have underlying spectra similar to
LBL BL~Lac type objects.  However three objects (the FR~I galaxy 3C264, 
and the FR~II galaxies 3C390.3  and 3C382) stand out as
having higher optical to submillimeter and X-ray to submillimeter luminosity
rations.  These three cores are candidate HBL or intermediate type
BL~Lac type objects.

Previous studies have found that the optical emission is enhanced
in some FR~II galaxies (e.g., \citealt{C02}), suggesting
that the optical emission in some cases arises from an additional component
such as an accretion disk.   The two FR~IIs with high optical
to submillimeter luminosity ratios in our sample also have high X-ray to submillimeter
ratios.  This suggests that a similar emission mechanism is responsible for
both excesses.  One explanation would be that some FR~II galaxies
have SEDs similar to HBLs rather than LBLs, in which case the emission
from the radio to the X-ray would be synchrotron emission.
We find only 1 FR~I with an HBL type SED whereas
we have found 2 such FR~IIs out of a smaller number of sources.
The larger number of FR~IIs is surprising since the unified model for BL~Lacs
by \citet{fossati,ghisellini} suggests that those objects with higher  luminosities
have lower energy synchrotron peaks,
whereas FR~II radio galaxies tend to have have higher luminosities 
than FR~I radio galaxies.  

Another possibility is that an additional emission mechanism is present
in both the optical and X-ray wavelengths in 3C390.3 and 3C382. These FR~IIs 
have been identified optically as broad line radio galaxies (BLRGs).
Recent work has found that the X-ray to radio ratios in BLRGs
quasars tend to be above
that expected from an extrapolation of lower energy objects \citep{HW99},
suggesting that there could be an additional component 
associated with an accretion disk accounting for 
the high optical to submillimter and X-ray to submillimeter flux ratios 
of 3C390.3 and 3C382.
FR~IIs with weak and strong emission lines may participate differently
in unified schemes; it may be predominantly weak-lined FR~IIs that
are the parent population of BL~Lac objects \citep{H98,willott}.

Recent work on BL~Lacs suggest that the SED depends on a beaming
factor as well as the energy density in the plasma \citep{ghi02}.
\citet{fossati,ghisellini,ghi02} found that higher luminosity BL~Lac type objects
tend to have lower energy synchrotron peaks. 
However, the objects considered here with likely synchrotron peaks
in the infrared have significantly lower power than 
the BL~Lac type objects studied by \citet{fossati}.
This makes it difficult to unify the 3CRR sample with the known
BL~Lac population, however as pointed out by \citet{C00, T03} if
the jets contain material accelerated at different velocities then it may be 
possible to resolve this problem.

The SEDs presented in this paper suffer from lack of infrared
coverage, diverse (non-uniform) types of measurements,
and are comprised of observations spanning many years.
Since these sources are variable, much of the scatter in the
points shown could be caused by intrinsic variability.  
It is likely 
that the shape of the SED depends on the current luminosity
of the source \citep{vagnetti}, an effect that would
be difficult to see given the scatter caused by the comparison
of points at different epochs.  Better wavelength
coverage, simultaneous observations and more uniform samples
would better allow better studies and tests of unification models.

\acknowledgements
We thank the HHT staff and the FOTs 
(Michael Dumke, Hal Butner, Paul Gensheimer)
for allowing these observations to be taken, helping us out
with observing, providing computer support and advice 
with the data reduction.  We also thank Shanna Shaked
and Heejong Seo for taking many of the observations.
We thank Bill Sparks, Meg Urry and Eliot Quataert for helpful discussions.

We acknowledge partial support from the REU program NSF grant PHY-0242483.
This work was in part supported by NASA through grant numbers
GO-07886.01-96A, and GO-07868.01-96A, 
from the Space Telescope Institute, which is operated by the Association
of Universities for Research in Astronomy, Incorporated, under NASA
contract NAS5-26555.
This research has made use of the NASA/IPAC Extragalactic Database
(NED) which is operated by the Jet Propulsion Laboratory,
California Institute of Technology, under contract with the
National Aeronautics and Space Administration.

\clearpage

\clearpage

\clearpage

\clearpage 

\begin{deluxetable}{lcc}
\tablecaption{$870\micron$ Bolometric observations}
\tablewidth{0pt}
\tablehead{
\multicolumn{1}{l}{Galaxy}     &
\colhead{ flux(mJy)} &
\colhead{ $ 2\sigma$ limit (mJy)} 
}
%
%
\startdata
     3C31 &      75.29 $\pm$      12.03 & \\ 
     3C35 &      -3.47 $\pm$      15.88 & $<$      31.75\\ 
     3C35 &     -10.75 $\pm$      12.89 & $<$      25.77\\ 
  3C66.0B &      91.36 $\pm$      12.43 & \\ 
     3C78 &     277.79 $\pm$      46.24 & \\ 
     3C84 &    1259.84 $\pm$      17.09 & \\ 
     3C88 &      71.49 $\pm$      15.66 & \\ 
     3C98 &     -19.79 $\pm$      12.10 & $<$      24.21\\ 
    3C111 &    4639.31 $\pm$      29.31 & \\ 
  3C136.1 &      -5.58 $\pm$      11.38 & $<$      22.75\\ 
    3C198 &      -0.66 $\pm$      17.00 & $<$      34.01\\ 
    3C198 &      -5.65 $\pm$      10.22 & $<$      20.45\\ 
    3C227 &     -31.88 $\pm$      20.40 & $<$      40.81\\ 
    3C227 &      11.62 $\pm$      10.32 & $<$      20.65\\ 
    3C236 &      47.93 $\pm$      10.14 & \\ 
    3C264 &      47.35 $\pm$      12.70 & \\ 
    3C270 &     166.71 $\pm$      18.41 & \\ 
  3C272.1 &     133.98 $\pm$      11.34 & \\ 
    3C274 &    1466.99 $\pm$      17.34 & \\ 
  3C277.3 &      15.63 $\pm$      10.63 & $<$      21.26\\ 
    3C293 &      47.88 $\pm$      15.47 & \\ 
    3C296 &      48.07 $\pm$      10.68 & \\ 
    3C305 &      28.57 $\pm$      13.28 & \\ 
    3C310 &      22.62 $\pm$      17.19 & $<$      34.38\\ 
    3C317 &      91.00 $\pm$      12.22 & \\ 
  3C318.1 &      -5.99 $\pm$       9.33 & $<$      18.66\\ 
    3C321 &      -2.40 $\pm$      12.84 & $<$      25.68\\ 
    3C326 &     -20.46 $\pm$      10.59 & $<$      21.18\\ 
    3C338 &      54.80 $\pm$      12.14 & \\ 
    3C353 &      73.76 $\pm$      10.11 & \\ 
    3C371 &     691.56 $\pm$      22.79 & \\ 
    3C382 &      61.97 $\pm$      11.93 & \\ 
  3C390.3 &     155.78 $\pm$      13.32 & \\ 
    3C402 &      26.55 $\pm$      13.13 & \\ 
    3C430 &      10.40 $\pm$      16.14 & $<$      32.28\\ 
    3C452 &      33.18 $\pm$      15.41 & \\ 
    3C465 &      76.20 $\pm$      13.54 & \\ 
  NGC6251 &     423.30 $\pm$      11.63 & \\ 
\enddata
\tablecomments{Some observations were repeated during better
conditions.  This set of observations was taken on the HHT
2001 Feb 10-12.  For the non-detections $2\sigma$ upper limits are
listed on the right. }
\end{deluxetable}

\begin{deluxetable}{lccccccccccc}
\tablewidth{0pt}
\tablecaption{Derived Spectral Indexes}
\tablehead{
\multicolumn{1}{l}{Galaxy}     &
\colhead{Luminosity@870$\mu$m} &
\colhead{$\alpha_{sr}$} &
\colhead{$\alpha_{os}$} &
\colhead{$\alpha_{xs}$} 
}
\startdata
%
      3C31.0   &   41.1   &  0.05     &    1.1    &    1.2    \\
     3C66.0B   &   41.4   &    0.16   &    1.0    &     1.1   \\
      3C78.0   &   42.2   &   0.29    &    0.9    &  \nodata  \\
      3C84.0   &   42.4   &    0.91   &    0.9    &    1.0    \\
      3C88.0   &   41.6   &   0.19    &     1.1   &  \nodata  \\
     3C111.0   &   43.8   &    -0.30  &    1.3    &  \nodata  \\
     3C236.0   &   42.5   &    0.13   &   \nodata  &  \nodata \\
     3C264.0   &    41.0  &   0.34    &      0.8   &   0.9    \\
     3C270.0   &   40.7   &   0.14    &    1.4     &   1.1    \\
     3C272.1   &   39.9   &   0.07    &     1.0    &   1.1    \\
     3C274.0   &   41.2   &    0.23   &     1.1    &   1.1    \\
     3C293.0   &   41.8   &   0.17    &  \nodata   & \nodata  \\
     3C296.0   &   41.2   &   0.11    &    1.3     &   1.0    \\
     3C305.0   &   41.5   &  0.01     &  \nodata   &   \nodata\\
     3C317.0   &   41.8   &   0.34    &     1.2    &   1.1    \\
     3C338.0   &   41.5   &   0.15    &    1.1     &    1.2   \\
     NGC6251   &   42.2   &   0.16    &    1.1     &   1.0    \\
     3C353.0   &   41.6   &   0.10    &  \nodata   & \nodata  \\
     3C371.0   &   43.0   &   0.22    &    0.8     &  \nodata \\
     3C382.0   &   42.1   &   0.26    &     0.4    &   0.7    \\
     3C390.3   &   42.5   &   0.16    &    0.7     &   0.8    \\
     3C402.0   &   41.0   & \nodata   &     0.9    & \nodata  \\
     3C452.0   &   42.1   &   0.32    & \nodata     &  \nodata\\
     3C465.0   &   41.6   &   0.30    &     1.0     &  1.0    \\
\enddata
\tablecomments{
Spectral indexes for sources detected at $870\mu$m.
The log of the power in erg/s, estimated from
the submillimeter flux, $\nu f_\nu 4 \pi D^2$ is also listed.
The distance $D$ is estimated using the redshift and a Hubble constant
of 75~km~s$^{-1}$~Mpc$^{-1}$.
$\alpha_{sr}$ is the spectral index between the 5GHz core and 
submillimeter ($870\mu$m).  The spectral indexes $\alpha_{os}$ and $\alpha_{xs}$
are those measured from  the optical and submillimeter
and x-ray and submillimeter flux densities, respectively.
X-ray fluxes  were taken from those compiled by 
\citet{HW99,HW00,S99,T03,C03,H01}, optical
nuclear fluxes from those compiled by \citet{C99,C00,C02b,HW00}.  
}
\end{deluxetable}

\begin{deluxetable}{lccccccccccc}
\tablewidth{0pt}
\tablecaption{Points used in Spectral Energy Distributions}
\tablehead{
\colhead{filter} &
\colhead{$\lambda$($\mu$m)} &
\colhead{flux (mJy)}  &
\colhead{DATEOBS}   &
\colhead{Ref.}   &
}
\startdata
\cutinhead{NGC 6251}
ROSAT/HRI &1.2e-3 &3.7e-4 & 1995 Jun 1  & HW00    \\
F547M   &0.547  &0.11    & 1996 Sep 13  & \nodata \\
F814W   &0.814  &0.20    & 1996 Sep 13  & \nodata \\
F110W   &1.10   &0.31    & 1998 Jun 7   & \nodata \\
F160W   &1.60   &0.54    & 1998 Jun 7   & \nodata \\
F205W   &2.05   &0.84    & 1998 Jun 7   & \nodata \\
LW1     &4.5    &$<5$    & 1996 Feb 26  & \nodata \\
LW2     &6.7    &9       & 1996 Feb 26  & \nodata \\
LW3     &14.3   &15      & 1996 Feb 26  & \nodata \\
345GHz  &870    &423     & 2001 Feb 12  & \nodata \\
10.7GHz &2.8e4  &600     & 1983 Mar     & J86  \\
5.0GHz  &6.0e4  &350     & 1983 Mar     & J86 \\
2.3GHz  &1.3e5  &250     & 1983 Mar     & J86 \\
1.7GHz  &1.8e5  &280     & 1983 Mar     & J86 \\
\cutinhead{3C31} 
Chandra & 1.2e-3 &1.1e-5   & 2000 Nov 6  & T03 \\
F702W   & 0.7    &  1.4    & 1995 Jan 19 & \nodata \\
LW2     & 6.7    &$<34$    & 1997 Jan 31 & \nodata \\
LW3     & 14.3   &$<26$    & 1997 Jan 31 & \nodata \\ 
345GHz  & 870    & 75      & 2001 Feb 12 & \nodata \\
5GHz    & 6e4    & 1000    & 1997 Feb 25 & G01 \\
1.7GHz  & 1.8e5  & 52      & 1997 Apr 9  & X00 \\
\cutinhead{3C66B} 
Chandra    & 1.2e-3 &3.0e-5 & 2000 Nov 20 & H01 \\
F25SRF2 & 0.253 & 9.6e-3  & 2000 Jul 13  & C02b \\
F814W   & 0.702 & 0.06    & 1999 Jan 31  & C02b \\
LW1     & 4.5   & $<5$    & 1997 Jan 26  & \nodata \\
LW2     & 6.7   & $<10$   & 1997 Jan 26  & \nodata\\
LW3     & 14.3  & $<13$   & 1997 Jan 26  & \nodata \\
345GHz  & 870   & 91      & 2001 Feb 10  & \nodata \\
5GHz    & 6e4   & 182     & 1993 Sep 12  & G01 \\
1.67GHz & 1.8e5 & 157     & 1997 Apr 9   & X00 \\
\cutinhead{3C78 (NGC 1218)}
BeppoSAX& 1.2e-3 &3.4e-5  & 1997 Jan 7   & T03 \\
F25QTZ  & 0.248 & 5.7e-2  & 2000 Mar 15  & C02b \\
F28X50LP& 0.722 & 0.66    & 2000 Mar 15  & C02b \\
345GHz  & 870   & 280     & 2001 Feb 11  &  \nodata \\
15GHz   &2e4    & 689     & 1982 Jun 18  & S86 \\
5GHz    &6e4    & 628     & 1982 Jun 18  & S86 \\
1.5GHz  &2e3    & 752     & 1982 Jun 18  & S86 \\
\cutinhead{3C84 (NGC 1275)}
ROSAT/HRI & 1.2e-3 &1.3e-3  &            & HW00  \\
F160W   & 1.6   & 4.3     & 1998 Mar 16  & \nodata \\
F702W   & 0.70  & $>1$    & 2000 Mar 3   & \nodata \\
667GHz  & 450   & 810     & 1998 Jul 15  & I01   \\
350GHz  & 850   & 1420    & 1998 Jul 15  & I01  \\
345GHz  & 870   & 1260    & 2001 Feb 11  & \nodata \\
15GHz   & 2e4   & 3000    & 1996         & D98 \\
\cutinhead{3C264}
ROSAT/PSPC & 1.2e-3 &4.9e-4  & 1991 Nov 29 & HW99    \\
F25CN182& 0.208 & 0.043   & 2000 Feb 12  & T03 \\
F547M   & 0.547 & 0.16    & 1996 May 19  & \nodata \\
F702W   & 0.702 & 0.20    & 1994 Dec 24  & \nodata \\
F110W   & 1.10  & 0.31    & 1998 May 12  & \nodata \\
F160W   & 1.60  & 0.44    & 1998 May 12  & \nodata \\
F205W   & 2.05  & 0.52    & 1998 May 12  & \nodata \\
345GHz  & 870   & 47.3    & 2001 Feb 11  & \nodata \\
5GHz    & 6e4   & 125     & 1993 Feb 25  & L97 \\
\cutinhead{3C270 (NGC 4261)}
Chandra & 2e-4  & 3.6e-5   & 2000 May 6  & C03 \\
F25SRF2& 0.253  & $<$2e-4 & 2000 Mar 5   & C02b \\
F547M &0.547   & 7.6e-3   & 1994 Dec 13  & \nodata  \\
F791M &0.791   & 1.8e-2   & 1994 Dec 13  & \nodata \\
F110W &1.10    & 3.8e-2   & 1998 Apr 23  & \nodata \\
F160W &1.60    & 0.11     & 1998 Apr 23  & \nodata \\
F205W &2.05    & 0.31     & 1998 Apr 23  & \nodata \\
L     &3.4     & 1.2      & 2000 Mar 21  & \nodata  \\
LW1   &4.5     & $<7$     & 1996 Jul 2   & \nodata \\
LW2   &6.7     & $<10$    & 1996 Jul 2   & \nodata \\
LW3   &14.3    & 19       & 1996 Jul 2   & \nodata  \\
345GHz&870     &167       & 2001 Feb 11  & \nodata  \\
43GHz &0.7e4   & 305      & 1997 Sep 7   & J00 \\
22GHz &1.4e4   & 165      & 1997 Sep 7   & J00 \\
8.4GHz&3.6e4   & 100      & 1995 Apr 1   & J97 \\
1.6GHz&1.87e5  & 100      & 1995 Apr 1   & J97 \\
\cutinhead{3C272.1 (M84,NGC 4374)}
Chandra  & 1.2e-3 & 4.4e-5   & 2000 May 19 & T03 \\
F547M    &  0.555 & 8.83e-2  & 1996 Mar 4  & B00 \\
F814W    &  0.814 & 1.38e-1  & 1996 Mar 4  & B00 \\
F110W    &  1.10  & 0.24     & 1998 Jul 13 & B00 \\
F160W    &  1.60  & 0.39     & 1998 Jul 13 & B00 \\
F205W    &  2.05  & 0.39     & 1998 Jul 13 & B00 \\
L        &  3.45  &$<2$      & 2000 Mar 22 & \nodata \\
LW2      &  6.7   &$<10$     & 1996 Jul 5  & \nodata \\
LW7      &  9.7   &$< 6$     & 1996 Jul 5  & \nodata \\
LW3      &  14.3  &$< 5$     & 1996 Jul 5  & \nodata \\
677GHz   & 450    &110       & 1999 Mar 19 & L00 \\
350GHz   & 850    &180       & 1999 Mar 19 & L00 \\
345GHz   & 870    &134       & 2001 Feb 10 & \nodata \\
221GHz   & 1350   &150       & 1999 Feb 14 & L00 \\
146GHz   & 2000   &150       & 1999 Mar 19 & L00 \\
8.09GHz  &3.71e4  & 190      & 1978        & J81 \\
1.67GHz  &1.80e5  & 160      & 1978        & J81 \\
\cutinhead{3C310}
ROSAT/HRI& 1.2e-3 &2.8e-5  & 1996 Jan 30  & HW99 \\
F25SRF2  & 0.253  & 3.41e-4& 2000 Jun 10  & C02b \\
F702W    & 0.70   & 5.7e-3 & 1994 Sep 12  & C02b \\
345GHz   & 870    &$ <34$  & 2001 Feb 10  & \nodata \\
5GHz     & 6e4    & 80     & 1973 May     & G88 \\
\cutinhead{3C317}
ROSAT/PSPC& 1.2e-3 &4.6e-5 & 2000 Sep 3   & T03 \\
F210M    & 0.221  & 7.7e-4 & 1994 Mar 5   & C02b \\
F702W    & 0.70   & 2.0e-2 & 1994 Mar 5   & C02b \\
F160W    & 1.6    & 0.12   & 1998         & T03 \\
345GHz   & 870    & 91     & 2001 Feb 10  & \nodata \\
5GHz     & 6e4     & 391    & 1989 Sep    & M93 \\
\cutinhead{3C338 (NGC 6166)}
Chandra  & 1.2e-3 &4e-6    & 1999 Dec 11  & T03  \\
F25SRF2  & 0.253  & 8.3e-4 & 2000 Jun 4   & C02b \\
F702W    & 0.70   & 1.6e-2 & 1994 Sep 9   & C02b \\
LW1      & 4.5    &$<3$    & 1996 Mar 2   & \nodata \\
LW2      & 6.7    &$<3$    & 1996 Mar 2   & \nodata \\
LW3      & 14.3   &$<3$    & 1996 Mar 2   & \nodata \\
345GHz   & 870    & 55     & 2001 Feb 11  & \nodata \\
8.4GHz   & 1e4    & 180    & 1990         & G98 \\
5GHz     & 6e4    & 105    & 1980 May 4   & G01 \\
1.4GHz   & 1e5    & 130    & 1990         & G98 \\
\cutinhead{3C382}
ROSAT/HRI& 1.2e-3   & 5.95e-3 & 1992 Mar 13  & HW99 \\
F25CN182 & 0.2078   & 2.3     & 2000 Feb 23  & C02b \\
F702W    & 0.700    & 8.0     & 1994 Jun 25  & C02b\\
LW2      & 6.7      & 50      & 1997 Feb 16  & \nodata \\
LW7      & 9.6      & 54      & 1996 Feb 16  & \nodata \\
345GHz   & 870      & 62      & 2001 Feb 11  & \nodata \\
5GHz     & 6e4      & 206     & 1990 Nov     & G01 \\
\cutinhead{3C390.3}
ROSAT/HRI& 1.2e-3 &4.3e-3  & 1995          & HW99 \\
F25CN182 & 0.208  & 2.9e-1 & 2000 Aug 10   & C02b \\
F702W    & 0.70   & 1.8    & 1994 Sep 20   & C02b  \\
345GHz   & 870    & 166    & 2001 Feb 12   & \nodata \\
5GHz     & 6e4    & 330    &               & G88    \\
\cutinhead{3C465}
ROSAT/HRI& 1.2e-3 &6.6e-5  & 1995 Jan 12  & HW99 \\
F25SRF2  & 0.253  & 4.5e-4 & 2000 May 25  & C02b \\
F702W    & 0.70   & 5.7e-2 & 2000 Jul 3   & C02\\
LW1      & 4.5    & $5.9$  & 1996 Dec 15  & \nodata \\
LW2      & 6.7    & $<4.7$ & 1996 Dec 15  & \nodata \\
LW3      & 14.3   & 2      & 1996 Dec 15  & \nodata \\
345GHz   & 870    & 76     & 2001 Feb 10  & \nodata \\
8.4GHz   & 3.6e4  & 146    & 1992 Jan     & V95 \\
5GHz     & 6e4    & 168    & 1992 Mar     & V95 \\
1.7GHz   & 2e4    & 181    & 1992 Jan     & V95 \\
\enddata
\tablecomments{
\baselineskip8pt
F25SRF2, F25QTZ, F25CN182, F28X50LP filters refer to STIS/HST UV 
and optical measurements.
The F110W, F160W, F205W filters refer to those from NICMOS (on board HST)
$1-2\mu$m.
F547M, F702W, F814W filters refer to optical WFPC2/HST measurements.
The LW1, LW2, LW3, LW7 filters refer to $4-15\mu$m data from ISOCAM images.
L band measurements are based on ground based images obtained by us 
at the IRTF.
The date of the observations are also listed.
}
\tablerefs{
\baselineskip8pt
B00 \citep{bower};
C02b \citep{C02b};
C03  \citep{C03};
D98 VLBI core \citep{dhawan};
G01 VLBI \citep{G01};
G98 \citet{G98};
G88 \citet{G88};
H01 \citep{H01};
HW00 mostly ROSAT \citep{HW00};
HW99 mostly ROSAT \citep{HW99};
I01 SCUBA \citep{irwin};
J81 VLBI \citep{jones81}
J86 VLBI \citep{jones86};
J97 VLBA \citep{jones97};
J00 VLBA \citep{jones00};
L00 SCUBA \citep{leeuw};
L97 VLBI \citep{lara97};
M93 \citep{M93};
S86 VLA and VLBI \citep{saikia};
T03 \citep{T03};
V95 VLBI \citep{venturi95};
X00 VLBA \citep{xu2000}.
When no reference is given we measured the fluxes ourselves using 
procedures described by \citet{almu}.
}
\end{deluxetable}

\end{document}